\documentclass[pra,aps,twocolumn,showpacs,superscriptaddress,floatfix]{revtex4}

\usepackage{amsmath,amssymb}
\usepackage{graphicx}
\usepackage{version}
\DeclareGraphicsExtensions{.eps,.ps}

\begin{document}

\title
{Number-of-Particle Fluctuations and Stability of Bose-Condensed
Systems}
\author{C.-H. Zhang}
\affiliation{Department of Physics, Indiana University, 727 E.\
3rd Street, Bloomington, IN 47405}
\date{\today}

\begin{abstract}

In this paper we show that a normal total number-of-particle
fluctuation can be  obtained consistently from the static
thermodynamic relation and dynamic compressibility sum rule. In
models using the broken $U(1)$ gauge symmetry, in order to keep the
consistency between statics and dynamics, it is important to
identify the equilibrium state of the system with which the density
response function is calculated, so that the condensate particle
number $N_0$, the number of thermal depletion particles $\tilde{N}$,
and the number of non-condensate particles $N_{nc}$ can be
unambiguously defined. We also show that the chemical potential
determined from the Hugenholtz-Pines theorem should be consistent
with that determined from the equilibrium equation of state. The
$N^{4/3}$ anomalous fluctuation of the number of non-condensate
particles is an intrinsic feature of the broken $U(1)$ gauge
symmetry. However, this anomalous fluctuation does not imply the
instability of the system. Using the random phase approximation,
which preserves the $U(1)$ gauge symmetry, such an anomalous
fluctuation of the number of non-condensate particles is completely
absent
\end{abstract}

\maketitle

\section{Introduction}

The number-of-particle fluctuation $\langle\delta N^2\rangle$ in an
equilibrium system is a fundamental statistic problem since its
scaling with the number of particles $\langle\delta N^2
\rangle\propto N^{\gamma}$ relates to the stability of the system.
The fluctuation is called normal if $\gamma=1$ and anomalous if
$\gamma>1$. In the latter case, it implies the system is unstable,
since the isothermal compressibility $\kappa_T\rightarrow\infty$ in
the thermodynamic limit (see Eq.\ (\ref{eq:fluct-kappa}) below). For
example, for a non-interacting uniform Bose system below the
critical temperature, the fluctuation of the number of condensate
particles $\langle\delta\hat{N}^2_0\rangle\propto N^2$, and that of
the number of non-condensate particles
$\langle\delta\hat{N}^2_{nc}\rangle\propto N^{4/3}$ in the grand
canonical ensemble, while
$\langle\delta\hat{N}^2_0\rangle=\langle\delta\hat{N}^2_{nc}\rangle
\propto N^{4/3}$ in the canonical ensemble\ \cite{fluct-IBG}; all
are anomalous since $\gamma>1$. However, for a trapped ideal Bose
gas, the fluctuation of the number particles is normal, since the
confinement effectively suppresses the thermal fluctuation\
\cite{trapped-IBG}.

Recently the number-of-particle fluctuation of interacting
Bose-condensed systems has attracted much theoretical attention, but
whether or not it is anomalous still has not been resolved, since
different methods predict different values
of $\gamma$\ \cite{Giorgini98,Illuminati99,Meier99,Idziaszek99,%
Xiong02,liu03,Pitaevskii03,Cherny05,Idziaszek05,Yukalov05}.
Particularly, even for the Bogoliubov approximation, both
$\gamma=4/3$ \ \cite{Giorgini98,Meier99,Xiong02,liu03,Idziaszek05}
and  $\gamma=1$\ \cite{Pitaevskii03,Yukalov05} scaling laws have
been obtained.

In order to see how these controversies  arise, it is useful to
examine the bases of these model calculations. Refs.\
\cite{Idziaszek99,Illuminati99} use an energy functional of the
total number of particles $N$, the number of thermal excited
particles $N_{ex}$, and the single-particle spectrum
$\varepsilon_{\vec{k}}$. Using this energy functional, the
fluctuations of the number of condensate and non-condensate
particles can be calculated using the partition function in either
the grand canonical ensemble, canonical ensemble, or microcanonical
ensemble, and a $\gamma=1$ scaling law was obtained for both the
condensate and non-condensate number-of-particles fluctuations. One
important observation, which is essential to obtain the $\gamma=1$
scaling law in this approach, is that phonon excitations have been
excluded from the single-particle spectrum. The $\gamma=1$ scaling
law for the condensate fluctuation is also obtained in Ref.\
\cite{Cherny05}, in which a single-condensate-mode
Hamiltonian is used. In Refs.\ \cite{Giorgini98,%
Xiong02,liu03,Idziaszek05}, the spectrum obtained by the Bogoliubov
approximation was used and a $\gamma=4/3$ scaling law was obtained
for the non-condensate number-of-particle fluctuation. However using
the compressibility sum rule (see Eq.\ (\ref{eq:compr-sum-rule})), a
$\gamma=1$ scaling law was obtained in Refs.\
\cite{Pitaevskii03,Yukalov05} for the total number-of-particle
fluctuation in the same Bogoliubov approximation. From this brief
literature survey, it is understood that the number-of-particle
fluctuation in an interacting Bose-condensate system is highly model
dependent, since the definition of the condensate fraction, the
number of non-condensate particles, and the energy spectrum are
highly model dependent. But the contradictory results in
Refs.\ \cite{Giorgini98,Xiong02,liu03,%
Idziaszek05} with that in Refs.\ \cite{Pitaevskii03,Yukalov05}
within the same Bogoliubov approximation deserve further
investigation.

As can be shown, the total number-of-particle fluctuations of any
equilibrium system can be calculated from the static thermodynamic
relation
\begin{align}
\label{eq:fluct-kappa} \frac{\langle\delta\hat{N}^2\rangle}{N}
=\frac{\langle\hat{N}^2\rangle-\langle\hat{N}\rangle^2}{N}
=\frac{k_BT}{N}\left.\frac{\partial
N}{\partial\mu}\right|_{T\Omega}=\rho k_BT\kappa_T,
\end{align}
where $\hat{N}$ is the particle number operator with expectation
value $N$, $k_BT$ is the temperature, $\mu$ is chemical potential,
$\Omega$ is the volume of the system, and $\rho=N/\Omega$ is the
number density. On the other hand, the total number-of-particle
fluctuation can be also determined by the following dynamic
compressibility sum rule
\begin{equation}
\label{eq:compr-sum-rule}
\frac{\langle\delta\hat{N}^2\rangle}{N}=-\frac{k_BT}{\rho}
\lim_{\vec{q}\rightarrow0}\chi_{nn}(\vec{q};\omega=0),
\end{equation}
where $\chi_{nn}(\vec{q};\omega)$ is the density response function.
The number-of-particle fluctuations calculated from these two
relations must be consistent in any approximation. However, we have
seen that the static result $\gamma=4/3$ obtained in Refs.\ \
\cite{Giorgini98,Xiong02,liu03,Cherny05} is not consistent with the
dynamic result $\gamma=1$ obtained in Ref.\
\cite{Pitaevskii03,Yukalov05}. This leads to a contradictory
conclusion about the stability of the system since, as argued by
Yukalov\ \cite{Yukalov05}, an anomalous fluctuation of the number of
non-condensate particles would inevitably lead the system to be
unstable while an interacting Bose-condensed system {\em is} stable.

We have seen that the above inconsistency of the statics with the
dynamics in the Bogoliubov approximation is related to the
separation of the condensate and non-condensate components when  the
broken Bose $U(1)$ gauge symmetry is used. In this case, the Bose
field operator $\hat{\psi}$ is split as
\begin{align}
\label{eq:Bogoliubov-scheme}
\hat{\psi}(\vec{r})=\Psi(\vec{r})+\delta\hat{\psi}(\vec{r}),
\end{align}
where $\Psi(\vec{r})\ne0$ is the Bogoliubov order parameter, and
$\delta\hat{\psi}(\vec{r})$, usually called the non-condensate field
operator,  represents both the dynamic excitation and thermal
depletion out of the condensate.  This subtlety in
$\delta\hat{\psi}$ shows that the condensate and non-condensate
components are strongly correlated, and the condensate component can
not be treated just as a static reservoir. Instead, the dynamics
aspect of $\Psi$ must be taken into account in calculation of the
number of condensate and non-condensate particles for the purpose of
calculating the number-of-particle fluctuation from Eq.\
(\ref{eq:fluct-kappa}).  To resolve the inconsistency of statics
with dynamics and to treat $\Psi$ as a dynamic quantity, it is
crucial to identify the equilibrium state with which the density
response function $\chi_{nn}$ is determined. Using this equilibrium
state as a reference, the number of condensation particles $N_0$,
the number of dynamically excited particles $N_{nc}$, and the number
of thermally depleting particle $\tilde{N}$ can be unambiguously
defined.

We will also show that the anomalous fluctuation of the number of
non-condensate particles is an intrinsic feature of the broken
$U(1)$ gauge symmetry. It is well known that a direct consequence of
the above broken $U(1)$ gauge symmetry is that the poles of the
single-particle Green function defined with $\delta\hat{\psi}$ and
the total density response function are identical. Therefore, the
fluctuation of the number of non-condensate particles inevitably
follows the $N^{4/3}$ anomalous law, since the momentum distribution
of this non-condensate particle always has a $1/k^2$ singularity in
the long-wavelength limit, regardless of at what level the
interacting Bose Hamiltonian is truncated. However, we shall show
that this anomalous fluctuation does not imply the instability of
the interacting Bose system. This is because, as we shall show, the
condensate and non-condensate components can not be treated as
linearly independent terms so that an anomalous fluctuation of the
number of non-condensate particles does not necessarily indicate
that the system is unstable.

One way to avoid such an anomalous fluctuation of the number of
non-condensate particles is to work with an ensemble in which the
gauge symmetry is {\it not} broken \cite{fix_n}, so that we can
avoid the entangling of particle-conserving collective excitations
and single-particle excited state in the pole of the non-condensate
single-particle Green function. Indeed, using the random phase
approximation with inclusion of exchange (RPAE) developed by
Minguzzi and Tosi\ \cite{Minguzzi97}, which keeps the $U(1)$ gauge
symmetry, we are able to show that while the total
number-of-particle fluctuation is normal and consistently determined
from statics and dynamics, the anomalous fluctuation of the number
of non-condensate particles is completely absent.

This paper is organized as follows. In Sec.\ \ref{sec:Bel-rules}, we
briefly summarize the rules to build the non-condensate
single-particle Green function and the density response function
with the broken $U(1)$ gauge symmetry. In Sec.\ \ref{sec:BA}, we
examine how the consistency between Eqs.\ (\ref{eq:fluct-kappa}) and
(\ref{eq:compr-sum-rule}) can be obtained in calculating the
number-of-particle fluctuation, and interpret the physical meaning
of the anomalous fluctuation of the number of non-condensate
particles in the Bogoliubov approximation. In Sec.\ \ref{sec:RPAE},
we carry out a calculation in the random phase approximation with
inclusion exchange and in the dielectric formalism to support our
interpretations presented in Sec.\ \ref{sec:BA}. The discussions and
conclusion are presented in the last section.

\section{Single-Particle Green function and Density
Response Function with Broken $U(1)$ Gauge Symmetry}
\label{sec:Bel-rules}

In this section, we briefly summarize the rules to construct the
single-particle Green function and density response function with
broken $U(1)$ gauge symmetry. The details can be found in Refs.\
\cite{Szep74,Fetter71,Griffin93}.

We start from the Hamiltonian for a homogenous interacting Bose
system in the second-quantized form
\begin{align}
\label{eq:Hamiltonian} \hat{H}=\sum_{\vec{k}}\mathcal{E}_{\vec{k}}
a^\dagger_{\vec{k}}a_{\vec{k}}+\frac{g}{2\Omega}
\sum_{\vec{q}\vec{k}_1\vec{k}_2}a^\dagger_{\vec{k}_1+\vec{q}}
a^\dagger_{\vec{k}_2-\vec{q}}a_{\vec{k}_2}a_{\vec{k}_1},
\end{align}
where a contact two-body potential with strength $g$ is used,
$\mathcal{E}_{\vec{k}}=\frac{\hbar^2\vec{k}^2}{2m}-\mu=\varepsilon_{\vec{k}}-\mu$
is the non-interacting single-particle energy with respect to the
chemical potential, and $a^\dagger_{\vec{k}}$ and $a_{\vec{k}}$ are
creation and annihilation operators for the interacting Bose
particles.

Now using Eq.\ (\ref{eq:Bogoliubov-scheme}) in its momentum space
form, i.e., replacing $a^\dagger_0$ and $a_0$ with $\sqrt{N_0}$,
where $N_0$ is the number of particles condensed onto the ground
state $\vec{k}=0$, one obtains the approximated Hamiltonian\
\cite{Griffin93}
\begin{align}
\label{eq:H-BT} \hat{H}&\approx\frac{gN_0^2}{2\Omega}-\mu
N_0+\sum_{\vec{k}\ne0}\mathcal{E}_{\vec{k}}
a^\dagger_{\vec{k}}a_{\vec{k}}+\frac{g\rho_0}{2}\sum_{\vec{q}\ne0}
\hat{A}_{\vec{q}}\hat{A}_{-\vec{q}} \nonumber\\
&+\frac{g\sqrt{N_0}}{2\Omega}
\left(\hat{A}_{\vec{q}}\hat{\tilde{\rho}}_{-\vec{q}}
+\hat{\tilde{\rho}}_{\vec{q}}\hat{A}_{-\vec{q}}\right) +
\frac{g}{2\Omega}\sum_{\vec{q}\ne0}
\hat{\tilde{\rho}}_{\vec{q}}\hat{\tilde{\rho}}_{-\vec{q}},
\end{align}
where $\rho=N_0/\Omega$ is the condensate density,
$\hat{A}_{\vec{q}}=a^\dagger_{\vec{q}}+a_{\vec{q}}$, and
\begin{align}
\hat{\tilde{\rho}}_{\vec{q}}=\sum_{\vec{k}\ne0,-\vec{q}}
a^\dagger_{\vec{k}}a_{\vec{k}+\vec{q}}
\end{align}
is the density operator for the non-condensate particles. The total
density operator is
\begin{align}
\hat{\rho}_{\vec{q}}=\sqrt{N_0}\hat{A}_{\vec{q}}+\hat{\tilde{\rho}}_{\vec{q}}.
\end{align}
Equation (\ref{eq:H-BT}) provides the starting point for many
approximations in which the broken $U(1)$ gauge symmetry is used.

The single-particle Green function matrix defined with
$\delta\hat{\psi}$ is \ \cite{Szep74,Griffin93}
\begin{align}
G_{\alpha\beta}(\vec{q}\ne0,\tau)=-\langle\mbox{T}_\tau
a_{\vec{q}\alpha}(\tau)a^\dagger_{\vec{q}\beta}\rangle,
\end{align}
where
\begin{align}
a_{\vec{q}\alpha}=\left\{\begin{array}{ll} a_{\vec{q}} \
\ \ \ \alpha=+,\\
a^\dagger_{-\vec{q}} \ \ \alpha=-\end{array}\right..
\end{align}
Solving the Dyson equation involving a $2\times2$ matrix self-energy
$\Sigma_{\alpha\beta}$, the single-particle Green function
$G_{\alpha\beta}$ has the general form\ \cite{Fetter71,Szep74}
\begin{align}
\label{eq:G-general} G_{\alpha\beta}(k)=\frac{(\alpha
i\omega_n+\mathcal{E}_{\vec{k}})\delta_{\alpha\beta}
+\alpha\beta\Sigma_{-\alpha-\beta}(k)} {D(k)},
\end{align}
where
\begin{align}
D(k)&=\left[i\omega_n-\mathcal{E}_{\vec{k}}-\Sigma_{++}
(k)\right] 
\left[i\omega_n+\mathcal{E}_{\vec{k}}+\Sigma_{--} (k)\right]
\nonumber\\& +\Sigma_{+-} (k)\Sigma_{-+} (k).
\end{align}
Here notation $k=(\vec{k};i\omega_n)$ is used. Various truncations
of the Hamiltonian (\ref{eq:H-BT}) correspond to select certain
types of self-energy diagrams in such a way that the
Hugenholtz-Pines theorem\ \cite{HPtheorem}
\begin{align}
\mu=\Sigma_{++}(0)-\Sigma_{+-}(0)
\end{align}
is satisfied, so that the pole determined by $D(k)=0$ is gapless in
the long-wavelength limit.

The density response function defined as
\begin{align}
\chi_{nn}(\vec{q};\tau)=-\langle\mbox{T}_\tau\hat{\rho}_{\vec{q}}(\tau)
\hat{\rho}_{-\vec{q}}(0)\rangle
\end{align}
can be written as\ \cite{Griffin93}
\begin{align}
\label{eq:chi-general}
\chi_{nn}(q)&=\sum_{\alpha\beta}\Lambda_{\alpha}
(q)G_{\alpha\beta}(q)\Lambda_\beta (q)
+\chi^R_{nn}(q),
\end{align}
where $\Lambda_\alpha$ is the vertex function describing process of
(de)excitations (in)out of the condensate and $\chi^R_{nn}$ is the
regular part response function. It is not obvious in this approach
to identify the equilibrium state with which the density response
function $\chi_{nn}$ is determined. We shall show in the next two
sections that the identification of such an equilibrium state is
important to unambiguously define the equilibrium condensate
particle number $N_0$, thermal depletion particle number
$\tilde{N}$, and the corresponding thermal depletion single-particle
Green function, which are used to build $\Lambda_{\alpha}$,
$\Sigma_{\alpha\beta}$, $G_{\alpha\beta}$, and $\chi^R$, and to
calculate the non-condensate particle number $N_{nc}$, so that the
consistency between statics and dynamics in calculating the
number-of-particle fluctuation can be obtained.

\section{Number-of-Particle Fluctuation in
Bogoliubov Approximation}
\label{sec:BA}

W now reexamine the number-of-particle fluctuation problem in the
Bogoliubov approximation at finite temperature. A comment is
deserved: even though we work at finite temperature, there are no
thermal depletion particles.

The vertex function is $\Lambda_\alpha=\sqrt{\rho_0}$, and the
self-energies are
\begin{align}
\Sigma_{++}(\vec{q};i\omega_n)&=\Sigma_{--}(\vec{q};i\omega_n)=2g\rho_0,\\
\Sigma_{+-}(\vec{q};i\omega_n)&=\Sigma_{-+}(\vec{q};i\omega_n)=g\rho_0,
\end{align}
and the chemical potential determined by Hugenholtz-Pines theorem is
\begin{align}
\label{eq:HP-BA} \mu=\Sigma_{++}(0;0)-\Sigma_{+-}(0;0)=g\rho_0.
\end{align}
Substituting the above $\Sigma_{\alpha\beta}$, $\Lambda_\alpha$ and
$\mu$ into Eq.\ (\ref{eq:G-general}), one gets the corresponding
single-particle Green functions for the non-condensate particles\
\cite{Fetter71,Griffin93}
\begin{align} \label{eq:G-BA}
G^{BA}_{++}(\vec{k};i\omega_n)&
=\frac{u^2_{\vec{k}}}{i\omega_n-\omega_{\vec{k}}}
-\frac{v^2_{\vec{k}}}{i\omega_n+\omega_{\vec{k}}},
\nonumber\\
G^{BA}_{+-}(\vec{k};i\omega_n)
&=-u_{\vec{k}}v_{\vec{k}} \left(
\frac{1}{i\omega_n-\omega_{\vec{k}}}-\frac{1}{i\omega_n+\omega_{\vec{k}}}\right),
\end{align}
where
\begin{align}
\omega^2_{\vec{k}}=\left[\varepsilon_{\vec{k}}-\Delta\right]
\left[\varepsilon_{\vec{k}}+2g\rho_0-\Delta\right],
\end{align} and
\begin{align}
u^2_{\vec{k}}&=\frac{\varepsilon_{\vec{k}}-\Delta+g\rho_0+\omega_{\vec{k}}}
{2\omega_{\vec{k}}},\\
v^2_{\vec{k}}&=\frac{\varepsilon_{\vec{k}}-\Delta+g\rho_0-\omega_{\vec{k}}}
{2\omega_{\vec{k}}}.
\end{align}
Here $\Delta=\mu-g\rho_0 $ has been introduced for future
convenience, and $\Delta=0$ for temperatures below $T_c$.

Substituting the vertex functions and the Green function into Eq.\
(\ref{eq:chi-general}), one gets the density response function
$\chi^{BA}_{nn}(\vec{q},i\omega_n)$ for the interacting Bose gas
\begin{align}
\label{eq:chi-BA}
\chi^{BA}_{nn}(\vec{q},i\omega_n)
=\frac{\rho_0\varepsilon_{\vec{q}}}
{\omega_{\vec{q}}}\left(\frac{1}{i\omega_n-\omega_{\vec{q}}}
-\frac{1}{i\omega_n+\omega_{\vec{q}}}\right).
\end{align}

We first calculate the number-of-particle fluctuation from dynamics.
Substituting Eq.\ (\ref{eq:chi-BA}) into Eq.\
(\ref{eq:compr-sum-rule}), one gets
\begin{align}
\label{eq:fluct-BA}
\frac{\langle\delta\hat{N}^2\rangle^{BA}}{N}=
\frac{k_BT}{mc_B^2},
\end{align}
where $c_B=\sqrt{g\rho_0/m}$ is the Bogoliubov phonon velocity.
Therefore, we get a normal number-of-particle fluctuation from the
dynamics.

Now let's use Eq.\ (\ref{eq:fluct-kappa}) to calculate  the
number-of-particle fluctuation. The total number of particles is
given
\begin{align}
\label{eq:n}
N=N_0+N_{nc},
\end{align}
where $N_{nc}$ is calculated as
\begin{align}
\label{eq:N-nc}
N_{nc}=-\frac{1}{\beta} \sum_{n,\vec{k}}
G^{BA}_{11}(\vec{k};i\omega_n)
=\sum_{\vec{k}\ne0}\left[(u_{\vec{k}}^2+v_{\vec{k}}^2)n_{\vec{k}}+
v_{\vec{k}}^2\right].
\end{align}
Direct calculation shows that
\begin{widetext}
\begin{align}
\label{eq:n-mu} \frac{k_BT}{N}\left.\frac{\partial
N_{nc}}{\partial\mu}\right|_{\mu=g\rho_0}=\frac{k_BT}{N}\left.\frac{\partial
N_{nc}}{\partial\Delta}\right|_{\Delta=0}
=\frac{1}{N}\sum_{\vec{k}\ne0}\left[(u_{\vec{k}}^2+v_{\vec{k}}^2)^2n_{\vec{k}}(n_{\vec{k}}+1)
+4u^2_{\vec{k}}v^2_{\vec{k}}\frac{k_BT}{\omega_{\vec{k}}}
\left(n_{\vec{k}}+\frac12\right)\right],
\end{align}
\end{widetext}
The anomalous $N^{4/3}$ behavior of this equation can be seen, since
in the long-wavelength limit, $u^2_{\vec{k}}\sim
v^2_{\vec{k}}\sim\frac{1}{k}$, and $n_{\vec{k}}\sim\frac{1}{k}$, so
the integrand has a $\frac{1}{k^4}$ singularity.

On the other hand, using the chemical potential $\mu=g\rho_0$, one
gets
\begin{align}
\label{eq:n0-mu} \frac{k_BT}{N}\left.\frac{\partial
N_0}{\partial\mu}\right|_{\mu=g\rho_0}=\frac{k_BT}{N}\left.\frac{\partial
N_0}{\partial\Delta}\right|_{\Delta=0}=
\frac{k_BT}{mc_B^2}.
\end{align}
The number-of-particle fluctuation is the sum of Eqs.
(\ref{eq:n-mu}) and (\ref{eq:n0-mu}), which is clearly not
consistent with Eq.\ (\ref{eq:fluct-BA}). This is the inconsistency
of the result in Refs.\ \cite{Giorgini98,Xiong02,liu03,Idziaszek05}
with that in Refs.\ \cite{Pitaevskii03,Yukalov05}. We should remind
ourselves that Eq.\ (\ref{eq:fluct-kappa}) is a thermodynamic
relation for an equilibrium system which is described by a set of
equations of state. Also, according to finite-temperature
linear-response theory, the density response function is calculated
from an equilibrium state. Of course the equilibrium states that are
used in Eqs.\ (\ref{eq:fluct-kappa}) and (\ref{eq:compr-sum-rule})
should be the same. So what is the equilibrium state for the
Bogoliubov approximation at finite temperature? To find the answer,
we notice that Eq.\ (\ref{eq:n0-mu}) is identical to Eq.\
(\ref{eq:fluct-BA}) and we get it from relation $\mu=g\rho_0$.
Therefore, the equilibrium state in the Bogoliubov approximation is
identified to be a state that all particles occupy in the
$\vec{k}=0$ level and its equation of state is given by
$\mu=g\rho_0=g\rho$. This identification is sound since the relation
$\mu=g\rho_0$ is exactly the time-independent Gross-Pitaevskii
equation for a uniform system without thermal depletion particles.
Indeed, as proved by Leggett\ \cite{Leggett01}, the Bogoliubov
Hamiltonian can be obtained from the Hamiltonian
(\ref{eq:Hamiltonian}) by keeping the terms that have non-zero
expectation values in a subclass of states built from
Gross-Pitaevskii ground state. If we adopt this identification:

(i) The number-of-particle fluctuation from statics is given by Eq.\
(\ref{eq:n0-mu}), which is now completely consistent with Eq.\
(\ref{eq:fluct-BA}). Both are normal, and therefore the system is
proved to be stable, as it should be.

(ii) $N_{nc}$ is the number of particles excited out of the
condensate due to its oscillation, i.e. it is the depletion of the
Gross-Pitaevskii equilibrium state\ \cite{Leggett01}. $N_0$ in Eq.\
(\ref{eq:n}) should be replaced by $N_0^\prime$,
\begin{align}
N^\prime_0=N_0-N_{nc},
\end{align}
the number of particles remaining in the condensate after $N_{nc}$
particles are dynamically excited out of the condensate.  The
single-particle Green function $G_{\alpha\beta}$ describes the
dynamic process of the oscillation of the condensate. With the
broken U(1) gauge symmetry, the oscillation of the condensate is
interpreted as ejecting particles, which become the non-condensate
particles. This interpretation of the single-particle Green's
function gives a reasonable explanation of the density response
function given by Eq.\ (\ref{eq:chi-BA}). It also explains the
physical meaning of the result given by Eq.\ (\ref{eq:n-mu}). We
argue that, the $N_{nc}$ dynamically excited particles form a
non-interacting system with chemical potential given by $\Delta=0$.
Equation (\ref{eq:n-mu}) then represents the number-of-particle
fluctuation of this non-interacting system. In order to see this, we
calculate the density response function of this non-interacting
system\ \cite{Griffin93}
\begin{widetext}
\begin{align}
\label{eq:chi-phonon}
\chi^{BA}_{nn,nc}(\vec{q};i\omega_n)=-\frac{1}{\Omega\beta}
\sum_{m\vec{k}\ne0}\left[G^{BA}_{++}(\vec{k};i\omega_m)G^{BA}_{++}(\vec{k}+\vec{q};
i\omega_m+i\omega_n)+G^{BA}_{-+}(\vec{k};i\omega_m)G^{BA}_{-+}(\vec{k}+\vec{q};
i\omega_m+i\omega_n)\right].
\end{align}
A similar result with Eq.\ (\ref{eq:chi-phonon}) is obtained by
Meier and Zwerger\ \cite{Meier99} by calculating the phase
fluctuation of the order parameter $\Psi(\vec{r})$ of Eq.\
(\ref{eq:Bogoliubov-scheme}). It is important to point out the
difference of the physical meanings between Eqs.\ (\ref{eq:chi-BA})
and (\ref{eq:chi-phonon}). According to Eq.\
(\ref{eq:compr-sum-rule}), the number-of-particle fluctuation of the
non-interacting system is
\begin{align}
\label{eq:s0} \frac{\langle\delta\hat{N}_{nc}^2\rangle}{N}=
-\frac{k_BT}{\rho}\lim_{\vec{q}\rightarrow0}
\chi^{BA}_{nn,nc}(\vec{q};0)=\frac{1}{N}
\sum_{\vec{k}\ne0}\left\{\left[\left(u^2_{\vec{k}}+v^2_{\vec{k}}
\right)^2+4u^2_{\vec{k}}v^2_{\vec{k}}\right]n_{\vec{k}}
\left(n_{\vec{k}}+1\right)+u^2_{\vec{k}}v^2_{\vec{k}}\right\},
\end{align}
\end{widetext}
where $n_{\vec{k}}=
(e^{\beta\omega_{\vec{k}}}-1)^{-1}$. Here
$\omega_{\vec{k}}=\sqrt{\varepsilon_{\vec{k}}(\varepsilon_{\vec{k}}
+2g\rho_0)}$ is the Bogoliubov mode. This is exactly the same as
Eq.\ (7) in Ref.\ \cite{Giorgini98} obtained by Giorgini {\em et.\!
al.} for the fluctuation of non-condensate particles $\langle \delta
N^2_{nc}\rangle/N$. We notice that the leading order terms of this
result is identical to those of Eq.\ (\ref{eq:n-mu}) in
$\vec{k}\sim0$ region, where the anomalous behavior arises, since
$\frac{k_BT}{\omega_{\vec{k}}}\approx n_{\vec{k}}$. Therefore, for
this non-interacting system, the number-of-particle fluctuations
obtained from statics and dynamics are also consistent, even though
they are anomalous. However, this anomalous fluctuation is not an
implication of instability of the interacting Bose gas, since we
have proved from both statics and dynamics that the total
number-of-particle fluctuation is normal. This can also be seen by
substituting Eq.\ (\ref{eq:n}) into Eq.\ (\ref{eq:fluct-kappa}) but
replacing $N_0$ with $N_0^\prime$; the anomalous fluctuation due to
$N_{nc}$ is completely canceled out. This calculation clearly shows
the importance of the dynamic aspect of the condensate reservoir.

(iii) There is a new consistency. The chemical potential as a
functional of the total number of particles $N$ and the equilibrium
number of particles $N_0$ in condensate can be determined both
dynamically from the Hugenholtz-Pines theorem (\ref{eq:HP-BA}) and
statically from the equilibrium equation of state. These two must be
consistent with each other. However, there is a deeper physical
meaning of this consistency. The equilibrium state described by the
equation of state has a definite number of particles (here is
$N_0$). Therefore, this new consistency shows that the
Hugenholtz-Pines theorem is to restore the conservation of the
number of particles. Indeed, it is well known that the
Hugenhotz-Pines theorem is required by the continuity equation\
\cite{Huang64,Hohenberg65}.

In the next section, we shall show that these interpretations about
the number-of-particle fluctuation, $N_0^\prime$, $N_{nc}$, and the
single-particle Green function $G_{\alpha\beta}$, as well as the
Hugenholtz-Pines theorem in the Bogoliubov approximation remain
valid at the level of approximation in which all the terms in Eq.\
(\ref{eq:H-BT}) are kept. As examples, we consider the random-phase
approximation with inclusion of exchange (RPAE) developed by
Minguzzi and Tosi\ \cite{Minguzzi97} and the dielectric formalism by
Fliesser {\em et. al.}\ \cite{Fliesser01}. We shall not give the
detailed derivations, since they are available in the literature.
The steps presented here highlight the physics at hand.

\section{Random-Phase Approximation and Dielectric
Approach with Inclusion Exchange}
\label{sec:RPAE}

In the RPAE, the equilibrium equations of state of the
Bose-condensed system are the time-independent finite-temperature
Gross-Pitaevskii equation for the condensate and static Hartree-Fock
equation for the thermal depletion particles. For a homogenous
system, they are
\begin{align}
\label{eq:GP-T}
\mu&=g\rho_{0}+2g\tilde{\rho},\\
\label{eq:hf-nc}
h_{HF}\psi_{\vec{k}}(\vec{r})&=\varepsilon^{HF}_{\vec{k}}
\psi_{\vec{k}}(\vec{r}),\ \; \vec{k}\ne0,
\end{align}
where
\begin{align}
\label{eq:hf-RPAE}
h_{HF}&=-\frac{\nabla^2}{2m}+2g\rho-\mu,\\
\varepsilon^{HF}_{\vec{k}}
&=\varepsilon_{\vec{k}}+2g\rho-\mu,
\end{align}
are the static Hartree-Fock Hamiltonian and single-particle energy
with respect to the chemical potential for the thermal depletion
particles. Here $\rho_{0}=N_{0}/\Omega$,
$\tilde{\rho}=\tilde{N}/\Omega$, and $\rho=N/\Omega$ are the
equilibrium condensate, thermal depletion, and total densities with
$N_{0}$, $\tilde{N}$ and $N$ the corresponding equilibrium
condensate, thermal depletion, and total number of particles. We
emphasize again that the number of particles in this equilibrium
system is conserved. By neglecting the thermal depletion
$\tilde{N}$, we arrive at the equilibrium equation of state for the
Bogoliubov approximation. We notice that single-particle orbits for
the condensate and thermal depletion particles are governed by two
different Hamiltonians, and therefore, they are not generally
orthogonal. However, for a uniform system, these single-particle
orbits are simple orthogonal plane waves. We also notice that there
is a gap in the single-particle spectrum
\begin{align}
\lim_{\vec{k}\rightarrow0}\varepsilon^{HF}_{\vec{k}}=g\rho_0.
\end{align}
This gap has important effects on many properties of a
Bose-condensed system\ \cite{Zhang05b}. We will come back this issue
in the last section.

In this paper, we concern its effect on the stability of the system,
as we shall show in the following.

We define a thermal depletion single-particle Green function for the
static $h_{HF}$
\begin{align}
\label{eq:G-depl}
\tilde{G}_{HF}(\vec{k};i\omega_n)=\frac{1}{i\omega_n-\varepsilon^{HF}_{\vec{k}}}.
\end{align}
The number of thermal depletion particles is found to be
\begin{align}
\label{eq:N-thermal}
\tilde{N}=-\frac{1}{\beta}\sum_{n,\vec{k}\ne0}\tilde{G}_{HF}(\vec{k};i\omega_n)
=\frac{\Omega}{\lambda_T^3}g_{3/2}(z),
\end{align}
where $\lambda_T=\sqrt{2\pi/mk_BT}$ and
$z=e^{\beta(\mu-2g\rho)}=e^{-\beta g\rho_0}$ and $g_{\gamma}(z)$ is
the Bose function. Equation (\ref{eq:N-thermal}) is the  equation of
state equivalent to Eq.\ (\ref{eq:hf-nc}) for the thermal depletion
particles. The self-consistent relations among $N$, $N_0$,
$\tilde{N}$ and $\mu$ are given by Eq.\ (\ref{eq:GP-T}) and
\begin{align}
\label{eq:N-equil}
N(\mu)=N_0(\mu)+\tilde{N}(\mu)=N_0(\mu)+\frac{\Omega}{\lambda_T^3}g_{3/2}[z(\mu)].
\end{align}

The number-of-particle fluctuation can be calculated by substituting
Eq.\ (\ref{eq:N-equil}) into Eq.\ (\ref{eq:fluct-kappa}). Using the
equations of states \ (\ref{eq:GP-T}) and (\ref{eq:N-thermal}), we
find
\begin{align}
\label{eq:dir-mu} \frac{\partial
N_0}{\partial\mu}&=-\frac{\Omega}{g}+2 \frac{\partial
N}{\partial\mu},
\nonumber\\
\frac{\partial\tilde{N}}{\partial\mu} &
=\frac{\beta}{\lambda^3_T}g_{1/2}(z)\left(\Omega
-2g\frac{\partial N}{\partial\mu}\right),
\end{align}
and as a consequence
\begin{align} \label{eq:fluct-N-s}
\frac{\langle\delta\hat{N}^2\rangle}{N} =k_BT\left.\frac{\partial N}
{\partial\mu}\right|_{T}
=\frac{\rho_0}{\rho}\frac{k_BT}{mc_B^2}\frac{1+g\tilde{P}_0}
{1+2g\tilde{P}_0},
\end{align}
where $\tilde{P}_0$ is defined as 
\begin{align}
\label{eq:P00}
\tilde{P}_0
=-\frac{\beta}{\lambda_T^3}g_{1/2}\left(e^{-\beta g\rho_0}\right).
\end{align}
Equation (\ref{eq:fluct-N-s}) reduces to Eq.\ (\ref{eq:n0-mu}),
obtained in Bogolibov approximation when one sets $\tilde{P}_0=0$
and $\rho_0=\rho$.

We must also point out that $k_BT\frac{\partial
N_0}{\partial\mu}\ne\langle\delta N_0^2\rangle$,
$k_BT\frac{\partial\tilde{N}}{\partial\mu}\ne\langle\delta
\tilde{N}^2\rangle$. Because of the ensemble used in the RPAE \
\cite{note-ensemble}, $\langle\delta N_0^2\rangle$ and
$\langle\delta \tilde{N}^2\rangle$ are actually
\begin{align}
\langle\delta N_0^2\rangle&\equiv0,\\
\label{eq:fluct-thermal}
\langle\delta\tilde{N}^2\rangle&=\sum_{\vec{k}}\tilde{n}_{\vec{k}}
(\tilde{n}_{\vec{k}}+1) =-z\frac{\partial\tilde{N}}{\partial
z}=\frac{\Omega\beta}{\lambda_T^3}g_{1/2}(z).
\end{align}
These two results are clearly not the same those given by Eq.\
(\ref{eq:dir-mu}). Therefore,
\begin{align}
\langle\delta N^2\rangle\ne\langle\delta
N_0^2\rangle+\langle\delta\tilde{N}^2\rangle.
\end{align}
This shows that, even in the mean-field level approximation, the
condensate and the thermal depletion components are strongly
correlated. This is not surprising, since below the critical
temperature, the presence of the condensate pins down the chemical
potential to be $\mu=2g\tilde{\rho}+g\rho_0$ and Eq.\
(\ref{eq:N-equil}) is a self-consistent relation between $N$ and
$\mu$. Even in the Bogoliubov approximation, it is this
self-consistent relation that predicts a number-of-particle
fluctuation given by Eq.\ (\ref{eq:n0-mu}) consistent with Eq.\
(\ref{eq:fluct-BA}) while the fluctuation of the condensate itself
is identically zero. Similar calculations for the RPA without the
exchange show that $\langle\delta\tilde{N}^2\rangle$ follows the
$N^{4/3}$ anomalous scaling law, but the total number-of-particle
fluctuation is
\begin{align} \label{eq:fluct-RPA}
\frac{\langle\delta\hat{N}^2\rangle}{N} =k_BT\left.\frac{\partial N}
{\partial\mu}\right|_{T} =\frac{k_BT}{mc_B^2},
\end{align}
which is normal.

We now calculate the density response function around the above
equilibrium state and calculate the number-of-particle fluctuation
from dynamics.

The linearized equations for the density fluctuation have the matrix
form\ \cite{Minguzzi97}
\begin{align}
\label{eq:chi-matrix}
\left(\begin{array}{c} \delta\rho_0 \\
\delta\tilde{\rho}\end{array}\right)=\left(\begin{array}{cc}
\chi_{cc} & \chi_{c\tilde{n}} \\
\chi_{\tilde{n}c} & \chi_{\tilde{n}\tilde{n}}\end{array}\right)
\left(\begin{array}{c} \delta U^c \\
\delta U^{\tilde{n}}\end{array}\right),
\end{align}
where $\delta U^c$ and $\delta U^{\tilde{n}}$ are the spatially and
time-varying  external potentials for the condensate and thermal
depletion components. We emphasize here that the matrix form
(\ref{eq:chi-matrix}) of the density response function is  simply a
result by splitting the system into a condensate and a thermal
depletion component, in which the number of particles is conserved,
instead of a result of using the broken $U(1)$ gauge symmetry\
(\ref{eq:Bogoliubov-scheme}) as claimed by Minguzzi and Tosi\
\cite{Minguzzi97}. The total density response function is then given
by
\begin{align}
\label{eq:chi-total}
\chi_{nn}=\chi_{cc}+\chi_{c\tilde{n}}+\chi_{\tilde{n}c}
+\chi_{\tilde{n}\tilde{n}}.
\end{align}

On the other hand, according to the linear response theory,
\begin{align}
\label{eq:delta-rho} \delta\rho_0&=\chi^0_c\delta
U^c_{HF}=\chi^0_c(\delta U^c+g\delta\rho_0+2g\delta\tilde{\rho}),
\nonumber\\
\delta\tilde{\rho}&=\chi^0_{\tilde{n}}U^{\tilde{n}}_{HF}
=\chi^0_{\tilde{n}}(\delta
U^{\tilde{n}}+2g\delta\rho_0+2g\delta\tilde{\rho}),
\end{align}
where $\chi^0_c$ and $\chi^0_{\tilde{n}}$ are the density response
functions of the condensate and the thermal depletion around the
equilibrium state, respectively. From the above two equations, the
four components are found
\begin{align}
\label{eq:chi-4} \chi_{cc}&=(1-2g\chi^0_{\tilde{n}})D^{-1}\chi^0_c,\
\ \chi_{\tilde{n}\tilde{n}}=(1-g\chi^0_c)D^{-1}\chi^0_{\tilde{n}},
\nonumber\\
\chi_{c\tilde{n}}&=2g\chi^0_cD^{-1}\chi^0_{\tilde{n}},\; \;
\;\;\;\;\;\;\;\;\;\;
\chi_{\tilde{n}c}=2g\chi^0_{\tilde{n}}D^{-1}\chi^0_c,
\end{align}
where
\begin{align}
D=(1-g\chi^0_c)(1-2g\chi^0_{\tilde{n}})-4g^2\chi^0_c\chi^0_{\tilde{n}}.
\end{align}

For homogenous systems, the density response functions of the
condensate and of the thermal depletion component  can be obtained
by linearizing the time-dependent Gross-Pitaevskii equation for the
condensate around Eq.\ (\ref{eq:GP-T}), and the time-dependent
Hartree-Fock equation for the non-condensate  around Eq.\
(\ref{eq:hf-nc}), respectively. They are given by
\begin{align}
\chi^0_c(\vec{q};i\omega_n)&=\frac{2\rho_0\varepsilon_{\vec{q}}}{(i\omega_n)^2
-\varepsilon^2_{\vec{q}}},\\
\chi^0_{\tilde{n}}(\vec{q};i\omega_n)&=\frac{1}{\Omega}\sum_{\vec{k}\ne0,-\vec{q}}
\frac{\tilde{n}_{\vec{q}+\vec{k}}-\tilde{n}_{\vec{k}}}
{i\omega_n+\varepsilon^{HF}_{\vec{q}+\vec{k}}
-\varepsilon^{HF}_{\vec{k}}}
\end{align}
where
$\tilde{n}_{\vec{k}}=(z^{-1}e^{\beta\varepsilon^{HF}_{\vec{k}}}-1)^{-1}$
is the occupation number of the static Hartree-Fock single-particle
level of the thermal depletion particles. Therefore, the total
density response function for the homogenous Bose-condensed system
is
\begin{widetext}
\begin{align}
\label{eq:chi-RPAE}
\chi_{nn}(\vec{q};i\omega_n)=\frac{\left[(i\omega_n)^2
-\varepsilon_{\vec{q}}^2\right]
\chi^0_{\tilde{n}}(\vec{q};i\omega_n)+2\rho\varepsilon_{\vec{q}}
\left[1+g\chi^0_{\tilde{n}}(\vec{q};i\omega_n)
\right]}{\left[(i\omega_n)^2-\varepsilon^2_{\vec{q}}\right]\left[1-
2g\chi^0_{\tilde{n}}(\vec{q};i\omega_n)\right]-2\rho\varepsilon_{\vec{q}}
\left[1+2g\chi^0_{\tilde{n}}(\vec{q};i\omega_n)\right]}.
\end{align}
\end{widetext}

Now substituting Eq.\ (\ref{eq:chi-RPAE}) into Eq.\
(\ref{eq:compr-sum-rule}), one gets the total number-of-particle
fluctuation from dynamics
\begin{align} \label{eq:fluct-N-d}
\frac{\langle\delta\hat{N}^2\rangle}{N}
=\frac{\rho_0}{\rho}\frac{k_BT}{mc_B^2}\frac{1+g\tilde{P}_0}
{1+2g\tilde{P}_0}.
\end{align}
Here we have made use of the fact that
$\lim_{|\vec{q}|\rightarrow0}\chi^0_{\tilde{n}}(\vec{q},0)
=-\frac{\beta}{\lambda_T^3}g_{1/2}\left(e^{-\beta
g\rho_0}\right)=\tilde{P}_0$ as given by Eq.\ (\ref{eq:P00}).
One can see that Eq.\ (\ref{eq:fluct-N-d}) is exactly the same as
Eq.\ (\ref{eq:fluct-N-s}).

We have shown the total number-of-particle fluctuation is normal and
consistent between statics and dynamics in the RPAE, and therefore,
the interacting Bose system is proved to be stable. In the above
derivations, the number of particles is conserved and there is not
any anomalous number-of-particle fluctuation. This is because in
this RPAE, the numbers of particles in the condensate and thermal
depletion component are not time-dependent\ \cite{Leggett01} and do
not change with the external potential because of entropy
conservation\ \cite{Blaizot86}. Therefore, they always take the
equilibrium values. The above results can be derived in a more
general time-dependent Hartree-Fock scheme which preserves the
$U(1)$ gauge symmetry\ \cite{Zhang05b}.

Now in order to see how the anomalous fluctuation arises when the
broken $U(1)$ gauge symmetry is used, we can follow the steps in
Ref.\ \cite{Fliesser01} to build the vertex function
$\Lambda_\alpha$, self-energy $\Sigma_{\alpha\beta}$, and the
single-particle Green function $G_{\alpha\beta}$ by using using
Eqs.\ (\ref{eq:GP-T}) and (\ref{eq:hf-nc}) as the reference. This
means that the equilibrium condensate $N_0$, thermal depletion
$\tilde{N}$, and Eq.\ (\ref{eq:G-depl}) should be used to build up
$\Lambda_\alpha$, $\Sigma_{\alpha\beta}$, and regular $\chi^R$, not
the as yet to be determined $N_0^\prime$, $G_{\alpha\beta}$, and
$N_{nc}$. Here we skip those steps and just cite the final results
for the single-particle Green function matrix below
\begin{widetext}
\begin{align}
\label{eq:G-dy}
G_{++}(\vec{q};i\omega_n)&=G_{--}(\vec{q};-i\omega_n)
=\frac{\left[i\omega_n-\varepsilon_{\vec{q}}\right]\left[1-
2g\chi^0_{\tilde{n}}(\vec{q};i\omega_n)\right]
+g\rho_0\left[1+g\chi^0_{\tilde{n}}(\vec{q};i\omega_n)
\right]}{\left[(i\omega_n)^2-\varepsilon_{\vec{q}}^2\right]\left[1-
2g\chi^0_{\tilde{n}}(\vec{q};i\omega_n)\right]
-2\rho_0\varepsilon_{\vec{q}}\left[1+2g\chi^0_{\tilde{n}}(\vec{q};i\omega_n)
\right]},
\displaybreak[0]\nonumber\\
G_{+-}(\vec{q};i\omega_n)&=G_{-+}(\vec{q};i\omega_n)
=-\frac{g\rho_0\left[1+g\chi^0_{\tilde{n}}(\vec{q};i\omega_n)
\right]}{\left[(i\omega_n)^2-\varepsilon^2_{\vec{q}}\right]
\left[1-2g\chi^0_{\tilde{n}}(\vec{q};i\omega_n)\right]
-2\rho_0\varepsilon_{\vec{q}}\left[1+2g\chi^0_{\tilde{n}}
(\vec{q};i\omega_n)\right]}.
\end{align}
\end{widetext}
and the density response function $\chi_{nn}(\vec{q};i\omega_n)$,
which is same as Eq.\ (\ref{eq:chi-RPAE}).

Since both the equilibrium state and the density response function
are the same as in RPAE, therefore, one gets the same consistent
normal number-of-particle fluctuations from statics and dynamics in
this dielectric formalism as those in the RPAE\ \cite{Fliesser01}.

The chemical potential from the Hugenholtz-Pines theorem in this
approximation turns out to be
\begin{align}
\label{eq:HP-RPAE}
\mu_{HP}=\Sigma_{++}(0;0)-\Sigma_{+-}(0;0)=g\rho_0+2g\tilde{\rho},
\end{align}
which is the exactly the same as Eq.\ (\ref{eq:GP-T}).

Using $G_{++}(\vec{q};i\omega_n)$, the number non-condensate
particles $N_{nc}$ and its fluctuation $\langle\delta
N_{nc}^2\rangle$ are found to be the same as Eqs.\ (\ref{eq:N-nc}),
(\ref{eq:chi-phonon}) and (\ref{eq:s0}). For a dilute gas $\rho
a^3\ll1$, where $a=\frac{gm}{4\pi}$, $\tilde{P}$ is usually very
small because of the single-particle gap\ \cite{Zhang05b}. It is
thus straightforward to show that the single-particle Green function
$G_{\alpha\beta}$ given by Eq.\ (\ref{eq:G-dy}) has the similar form
as that in Bogoliubov approximation for small $\vec{k}$. For
example, the pole is given by $\omega_{\vec{k}}\approx
c_Bk\left[1+2g\tilde{P}(\vec{k},\omega=c_Bk)\right]$. Therefore,
following the steps as in the Bogoliubov approximation, one can show
that $\langle\delta N_{nc}\rangle$ follows the $\gamma=4/3$ scaling
law.

Since $\tilde{N}$ from Eq.\ (\ref{eq:N-thermal}) is the number of
thermal depletion particles, the difference
\begin{align}
\delta N_0=N_{nc}-\tilde{N}
\end{align}
can be interpreted as the number of particles excited out of the
condensate by the oscillation of the whole system induced by the
external time-dependent potential. Indeed, in the Bogoliubov
approximation $\delta N_0=N_{nc}$ since the depletion of the
condensate is completely caused by the dynamic collective
excitation. Therefore, the single-particle Green functions
(\ref{eq:G-dy}) can be interpreted as dynamic ones comparing to the
thermal depletion $\tilde{G}_{HF}$ defined by Eq.\
(\ref{eq:G-depl}). As in Bogoliubov approximation, this
interpretation is allowed only because of the broken $U(1)$ gauge
symmetry. The total number of particles is expressed as
\begin{align}
\label{eq:N-d} N=N^\prime_0+N_{nc},
\end{align}
where $N_0^\prime=N_0-\delta N_0$. When substituting Eq.\
(\ref{eq:N-d}) into Eq. (\ref{eq:fluct-kappa}) to calculate the
number-of-particle fluctuation, the anomalous fluctuation due to
$N_{nc}$ is exactly canceled out by the one from $N^\prime_0$, so
that the total number-of-particle fluctuation is normal, which is
given by Eq.\ (\ref{eq:fluct-N-s}).

We can see that the anomalous fluctuation of the number of
non-condensate particles $\langle\delta N_{nc}^2\rangle$ is solely
due to the single-particle Green functions defined by Eq.\
(\ref{eq:G-general}) whose poles entangle the single-particle and
particle-conserving collective excitations, a directly consequence
of the $U(1)$ symmetry breaking rather than an implication of the
instability of the Bose system since the total number-of-particle
fluctuation is normal and consistent from statics and dynamics. More
than thirty years ago, Straley advised caution\ \cite{Straley72} in
using such a single-particle Green function to describe the
zero-sound characteristic spectrum of the superfluid $^4$He. Leggett
also argued\ \cite{Leggett01} that there are no circumstances in
which Eq.\ (\ref{eq:Bogoliubov-scheme}) is physically correct.

\section{Discussions and Conclusion}
\label{sec:conclusion}

We have shown that the anomalous fluctuation of the number of
non-condensate particles is an intrinsic feature of the broken
$U(1)$ gauge symmetry and is completely absent in the RPAE in which
the $U(1)$ gauge symmetry is preserved. This may be just related to
the different interpretations of the dynamic process of the
condensate of these models, as we point out in previous sections.
But since this anomalous fluctuation of the number of non-condensate
particles has not any physical significance, we can safely say that
it is just a by-product of using the broken $U(1)$ gauge symmetry.

An advantage to keep the $U(1)$ gauge symmetry is that the
single-particle spectrum and the collective excitation spectrum are
distinguished. As shown in the RPAE, they are the poles of thermal
depletion single-particle Green function and the density response
function, respectively. But single-particle spectrum has important
effects on the collective mode. For example, because of the gap,
$\chi^0_{\tilde{n}}$ is quite small for a dilute gas so that the
thermal depletion component does not sustain a zero sound mode. The
whole effect of the thermal depletion particles is to shift the
Bogoliubov mode. Also the gap causes the damping of the thermal
depletion particles to the collective mode exponentially decreases
when the temperature drops. As we recently discussed\
\cite{Zhang05b}, this single-particle gap is also responsible for
the completely screening of the external potential by the
condensate.

By using the broken $U(1)$ gauge symmetry, the boundary of the
single-particle spectrum and the collective mode are entangled
together if the poles of $G_{\alpha\beta}$ are interpreted as the
single-particle excitations. But as far as the dielectric formalism
in Ref.\ \cite{Fliesser01} concerned, there is a gapped
single-particle spectrum as same as that in the RPAE for the
equilibrium reference. In this sense, a gapped single-particle
spectrum and a gapless collective mode do coexist even in the
dielectric formalism. This is another point to identify the
equilibrium state with which the density response function is
calculated.

As pointed out by Meier and Zwerger\ \cite{Meier99}, the anomalous
fluctuation of the non-condensate particles is related to the
gapless mode in the superfluid Bose system. Our analysis shows this
is the case if one  uses the broken Bose $U(1)$ gauge symmetry. In
the RPAE, which preserves this gauge symmetry, there is no such an
anomalous number-of-particle fluctuation related to the gapless
superfluid mode.

In conclusion, we have shown that in models using the broken $U(1)$
gauge symmetry, the number-of-particle fluctuation is normal and can
be calculated consistently from the static thermodynamic relation
and dynamic compressibility sum rule if the equilibrium states are
identified. We also show that the chemical potential determined from
the Hugenholtz-Pines theorem should also be consistent with that
determined from the equilibrium equation of state. The $N^{4/3}$
anomalous fluctuation of the number of non-condensate particles is
an intrinsic feature of the broken $U(1)$ gauge symmetry. However,
this anomalous fluctuation does not imply the instability of the
system. Using the RPAE, which preserves the $U(1)$ gauge symmetry,
such an anomalous fluctuation of the number of non-condensate
particles is completely absent.

\acknowledgements

This work was supported by the NSF Grant No. DMR0454699. The author
thanks H. A. Fertig  and V. I. Yukalov for helpful discussions which
inspired this work. The author also thanks David Cardamone for
reading the manuscript.

\bibliography{bec-zf1}

\end{document}